\documentclass[twocolumn,10pt,a4paper]{article}

\usepackage[utf8]{inputenc}
\usepackage{graphicx}
\usepackage{amsmath,amssymb}
\usepackage{hyperref}
\usepackage{geometry}
\usepackage[normalem]{ulem}
\title{\vspace{-1.5cm} \textbf{The Preprint Evolution: The Rise of Unreviewed Drafts on \texttt{arXiv} and its Implications for Astronomy}}
\author{
  \textbf{Luigi Rolly BEDIN}\\
  \small \textit{Istituto Nazionale di Astrofisica, Osservatorio
    Astronomico di Padova, Vicolo dell'Osservatorio 5, Padova I-35122,
    Italy}\\
}
\date{} 

\begin{document}

\twocolumn[
  \begin{@twocolumnfalse}
    \maketitle
%
%
\begin{abstract}
\noindent The \texttt{arXiv} preprint server originally democratized
astronomical research by
removing the financial barriers of journal subscriptions. However, 
a system designed for the lower publication volumes of the 1990s has 
fractured under modern research output.
Today, the sheer volume of unreviewed manuscripts---often tagged as
\textit{``submitted,''}, 
\textit{``under review''},  
or
\textit{``comments welcome''}---imposes
a severe cognitive ``tax'' on the community.
Researchers are forced to track multiple, unvetted versions of the
same work, creating a massive redundant workload. This practice also
bypasses the blinded peer-review process by establishing premature
precedence, and it takes a heavy toll on early-career researchers.
The pressure to rapidly post unvetted results stifles independent
research, pushing young scientists toward massive mainstream
collaborations where individual creativity gets lost.
In this Short Communication, I analyze a decade of \texttt{astro-ph}
metadata to quantify this trend. To safeguard both researchers' time
and research freedom, I propose
%
%
bifurcating submissions into two separate streams.
\texttt{Level\,1} (\textit{peer-reviewed science}) would host peer-reviewed science (accompanied by a
journal acceptance or final DOI), while
\texttt{Level\,2} (\textit{pre-review drafts}) would be reserved for discussion drafts. This
distinction would remove the unfair advantage of establishing false
precedence through drafts and drastically cut the daily reading burden
by allowing researchers to opt out of tracking unvetted versions.
%
\end{abstract}

%
    \vspace{0.5cm} 
  \end{@twocolumnfalse}
]


\section{Introduction}
The \texttt{astro-ph} repository was built on a straightforward
premise: global, immediate access to scientific progress, free from
geographic or financial barriers. It sought to digitize the historical 
practice of mailing physical preprints to colleagues, providing a fast 
and free way to share early claims. However, a model designed for the 
occasional circulation of early drafts has become unsustainable under 
the sheer weight of modern astronomical output. A look at today's daily
\texttt{arXiv} mailing list reveals that a huge fraction of the roughly 100 
daily postings are unvetted drafts carrying disclaimers like 
\textit{``submitted''},
\textit{``to be submitted''} or
\textit{``comments are welcome''}. 

Scaling this preprint culture to current publication volumes has real
consequences.  The most obvious is the multiplicative
workload. Astronomical literature is expanding rapidly on its own. By
normalizing the daily upload of dozens of unreviewed drafts---and
their subsequent revised versions---the community is forced to read
the same manuscript multiple times. A researcher must read a
preliminary version today, and then re-evaluate the final paper months
later just to figure out what actually survived peer review. This
redundant tracking places an unsustainable burden on everyone's time,
turning the daily \texttt{arXiv} check into a grueling filtering
exercise.

Beyond the wasted time, this practice compromises the integrity of
peer review. Posting unreviewed manuscripts allows authors (often
senior scientists with large platforms) to prematurely plant a flag on
specific ideas. This public staking of claims can influence anonymous
referees, bypassing the editor's role in ensuring an impartial
evaluation.

\section{The Impact on Research Freedom and Early-Career Scientists}
Historically, paradigm shifts and breakthroughs come from individual
minds or small, highly focused teams. Innovation requires the freedom
to think independently, make private mistakes, and refine ideas before
exposing them to public scrutiny.

The current \textit{``publish first, review later''} culture destroys this
incubator. Young scientists, seeing the need to maintain constant
visibility on the daily feed, feel pressured into a frantic pace of
output. To survive in a volume-driven environment, early-career
researchers are increasingly pushed to abandon small, independent
projects and instead join mega-collaborations numbering in the
hundreds or thousands.

While massive collaborations are essential for large-scale
observational cosmology or instrumentation, funneling an entire
generation into them is a mistake. It breeds a factory-like approach
to science, channeling intellects into mainstream workflows where
individual identity disappears. The right to research freedom---the
space to independently explore a niche without the anxiety of
immediate, unreviewed dissemination---is slowly eroding.

\section{The Evolution of \texttt{astro-ph} Postings: A Statistical Analysis}
While the NASA Astrophysics Data System
(ADS)\footnote{\url{https://ui.adsabs.harvard.edu}} remains the
definitive tool for tracking finalized, peer-reviewed astronomical
literature, it does not inherently capture the transient,
self-reported submission statuses of preprints.
Therefore, to quantify the shift from accepted literature to
preliminary drafts within the daily feed, I utilized the
\texttt{arXiv} \textit{Application Programming Interface} (API).

The \texttt{arXiv}
API\footnote{\url{https://info.arxiv.org/help/api/}} is a public web
service that allows researchers to programmatically search and extract
metadata from the repository without manual web scraping. It is
accessed by sending HTTP requests to a dedicated endpoint
(\url{http://export.arxiv.org/api/query}), which returns structured
data in an Atom/XML format.

A Python script systematically queried this endpoint,
specifically targeting the \texttt{cat:astro-ph*} category, to extract
the metadata for all submissions between January 2016 and August
2026. By parsing the XML response for each paper, I isolated the
\texttt{arxiv:comment} field, which contains the authors'
self-declared publication status. The data extraction was finalized on
August 28, 2026.
%
I categorized the papers into three levels based on the
self-reported
\texttt{``Comments''} field:
\textbf{Accepted}   (e.g., \textit{``accepted for publication''}, \textit{``in press''}), 
\textbf{Pre-review} (e.g., \textit{``submitted''}, \textit{``in prep,''}, \textit{``comments welcome''}), and
\textbf{Unspecified} (no status comment provided).

\begin{figure}[h]
    \centering
    \includegraphics[width=\linewidth]{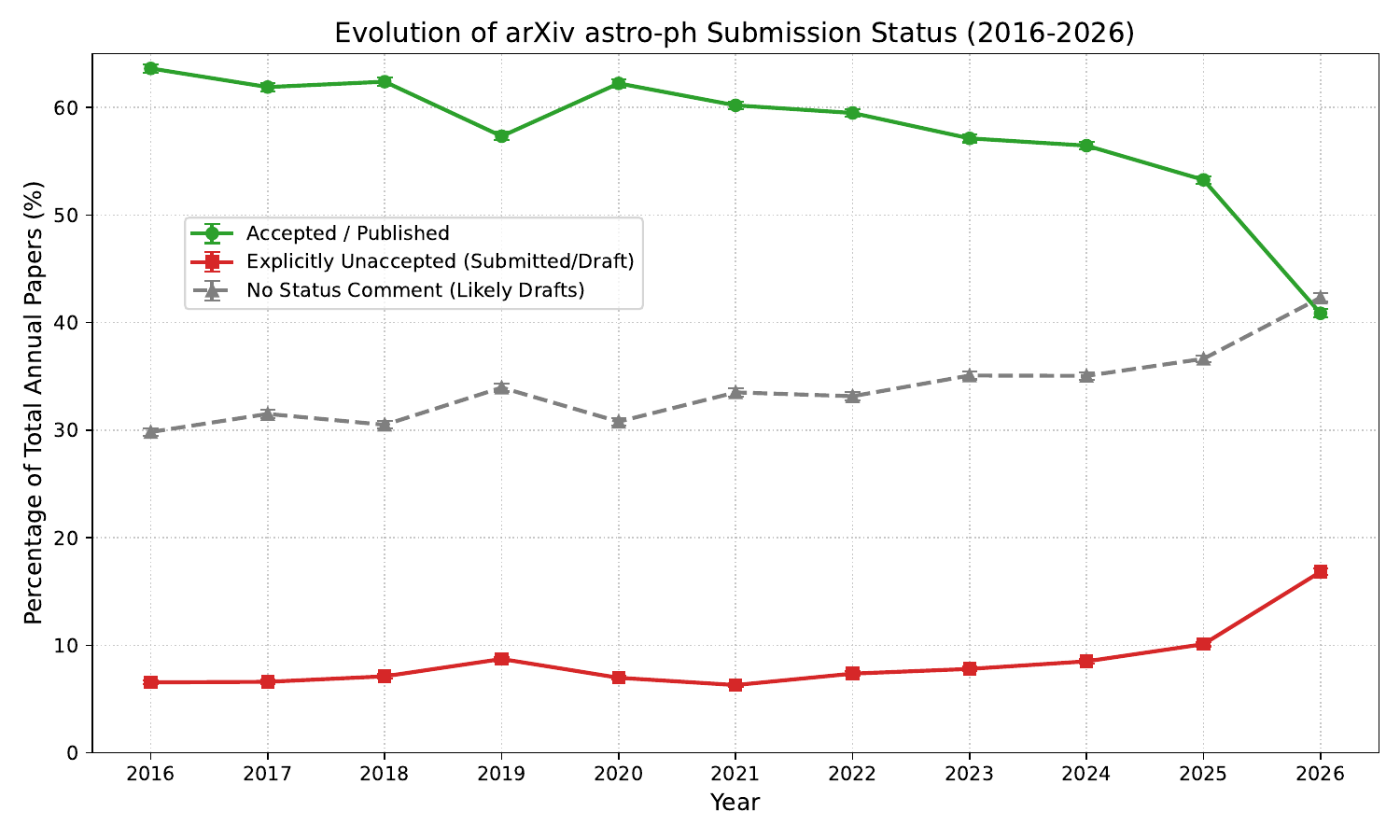}
    \caption{Temporal evolution of the submission status on
      \texttt{arXiv astro-ph} (2016-2026). The steady decline
      in explicitly accepted papers is mirrored by a massive baseline
      of unaccepted drafts and unclassified submissions.
      Data updated
      as of August 28, 2026.}
    \label{fig:trend}
\end{figure}

The data reveals a dramatic shift in publication habits
(Figure~\ref{fig:trend}). It is crucial to note that querying the
\texttt{arXiv} API retrieves the \textit{current} metadata of a
manuscript. Thus, historical data (2016--2025) represents the
``finalized'' state of papers after authors had months or years to
update their drafts to accepted status. Even with this long-term
correction, the data is alarming. In 2016, out of 15,179 total
submissions, 36.4\% remained permanently unvetted (either explicitly
marked as drafts or lacking any status). By 2025, as total annual
submissions surged to nearly 21,900, this fraction of permanently
uncertified literature had climbed to almost 47\%.\footnote{
      It is interesting 
      to note the anomalous data point in 2019, coinciding with the onset 
      of the COVID-19 pandemic. 
      }

The 2026 dataset (covering up to August) provides an even more
critical insight. While this partial year is inherently subject to
seasonal submission effects and its recent uploads have not yet had
time to clear the lengthy astronomical peer-review process, it acts as
a highly accurate, real-time snapshot of the daily \textit{astro-ph}
feed (the ``v1'' experience). Here, explicitly accepted papers
(40.9\%) are definitively eclipsed by manuscripts lacking a status
comment (42.3\%) and explicit drafts (16.8\%). Combined, nearly 60\%
of the current daily influx consists of unvetted
works-in-progress. This perfectly quantifies the modern researcher's
daily experience: \texttt{arXiv} is no longer an archive of finalized
science, but overwhelmingly a bulletin board of \textit{preliminary claims}.

\section{A Proposal for Structural Reform}
I am not advocating for \textit{any} restriction on open science; rather, I
propose restoring order to it. Platforms like \texttt{arXiv} can
maintain the democratization of research while mitigating the damage
of unvetted dissemination through a simple structural change: a
dual-level repository architecture.

The \texttt{arXiv} daily feed and web interface should clearly bifurcate submissions into two strictly separated streams:
\begin{itemize}
    \item \textbf{\texttt{Level\,1:} Peer-Reviewed Science.} Manuscripts accompanied by a journal acceptance or final DOI.
    \item \textbf{\texttt{Level\,2:} Discussion Drafts.} Manuscripts submitted for review or seeking community feedback.
\end{itemize}

The core feature must be the ability for users to completely filter
out Level 2 from their daily alerts. This distinction would remove the
unfair advantage of establishing false precedence through drafts.
It would drastically cut the daily reading burden by allowing
researchers to opt out of tracking unvetted versions.
Finally, since costly Open Access models are not yet a universal
standard, it would ensure that researchers in developing nations---who
rely heavily on \texttt{arXiv} as their primary library---retain
direct access to finalized, reliable science.

\section{Conclusions}
True democratization of science means universal access to verified
knowledge, not equal exposure for every rough draft. By treating our
primary repository as a workspace rather than an archive, the
astronomical community is fostering an environment that stifles
independent thought, compromises peer review, and herds our brightest
young minds into monolithic research factories.
Implementing a clear line between accepted papers and unreviewed
drafts is a necessary step to protect research freedom, secure the
future of astronomical innovation, and guarantee massive time savings
for researchers who opt out of tracking unvetted literature.

\section*{Acknowledgements}
%
The author thanks the anonymous referee for the prompt review and the
useful suggestion, as well as the
\textit{Astronomische Nachrichten} editorial office for their speed
and efficiency in handling this manuscript.

~\\

\newpage

\begin{figure}[h]
    \centering
    \includegraphics[width=\linewidth]{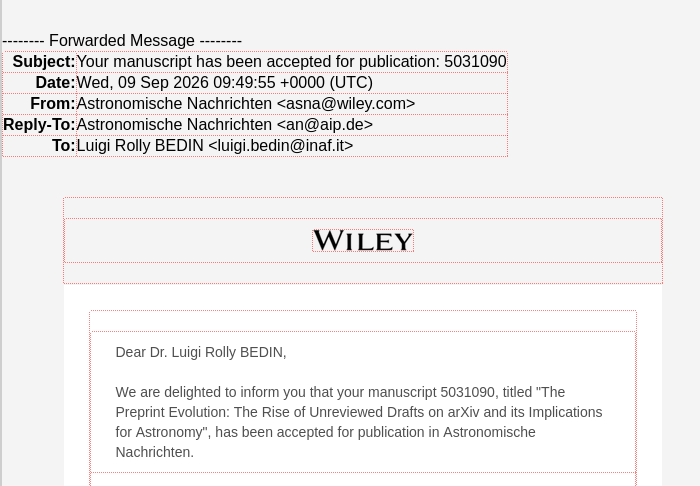}
    \caption{Official acceptance notification from Astronomische Nachrichten, received on September 9, 2026.}
    \label{fig:Acc}
\end{figure}

\begin{figure}[h]
    \centering
    \includegraphics[width=\linewidth]{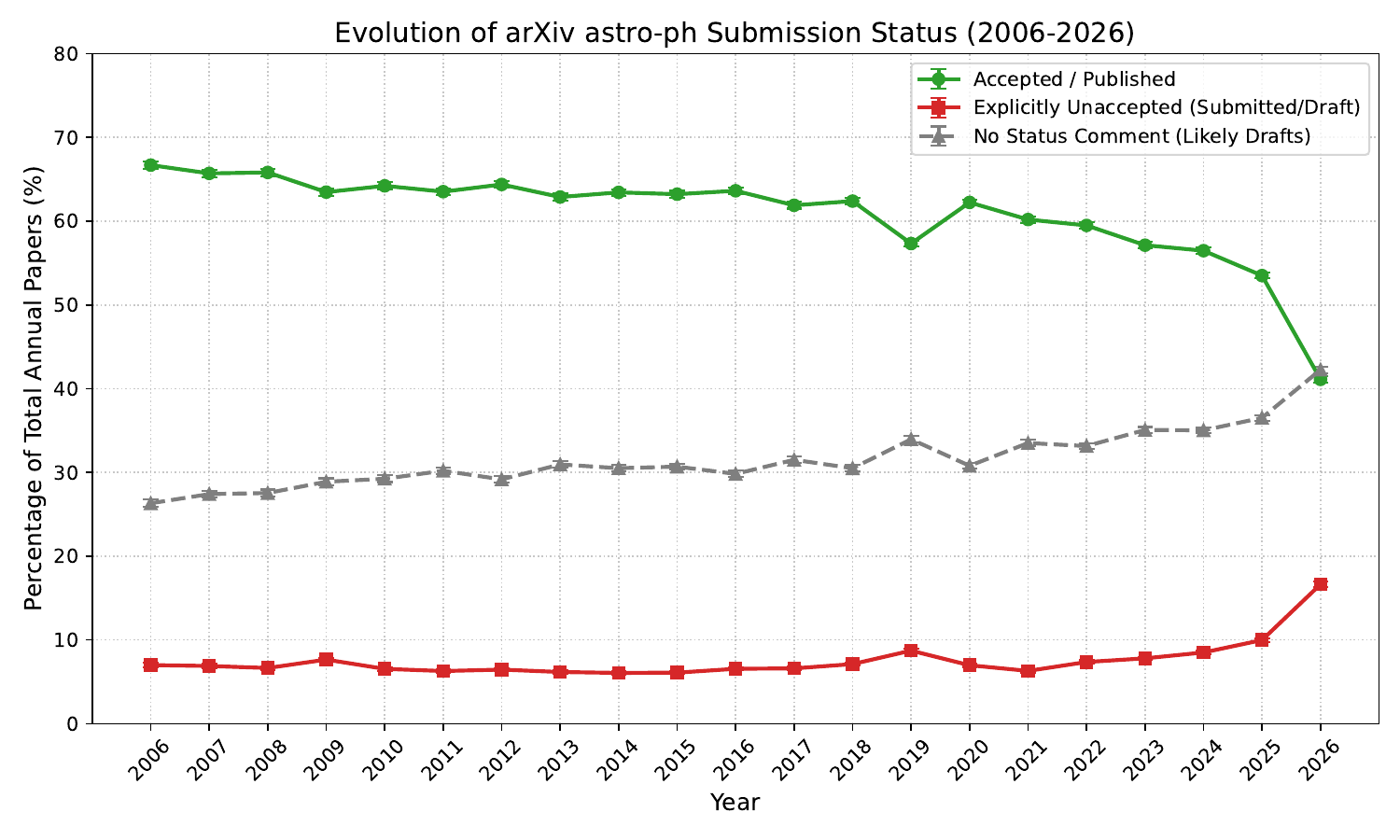}
    \caption{\textbf{Bonus material:} Same as Fig.\,\ref{fig:trend}, but from 2006 to 2026.}
    \label{fig:trend20yrs}
\end{figure}

\noindent
\texttt{Manuscript ID: 5031090}\\
\texttt{Article ID: ASNA70138}\\
\texttt{Article DOI: 10.1002/asna.70138}\\
\texttt{Internal Article ID: 100686201}\\
\texttt{Journal: Astronomische Nachrichten}\\
\texttt{Submitted: September 1, 2026}\\
\texttt{Accepted: September 9, 2026}\\


\begin{thebibliography}{1}

\bibitem{bedin2026}
Bedin, L.\,R. (2026). 
The Preprint Evolution: The Rise of Unreviewed Drafts on arXiv and its Implications for Astronomy. 
\textit{Astronomische Nachrichten}. 
Article DOI: \href{https://doi.org/10.1002/asna.70138}{10.1002/asna.70138} 
(Submitted: September 1, 2026 | Accepted: September 9, 2026). 
Manuscript ID: 5031090.

\end{thebibliography}
\end{document}